\documentclass[twocolumn,aps,pre]{revtex4}
\usepackage[dvips]{graphicx}
\usepackage{amsfonts}
\usepackage{epsfig}
\usepackage{amsmath}
\usepackage{amssymb}

\begin{document}

\title{Self diffusion in a system of interacting Langevin particles} 

\author{D. S. Dean$^{1,2}$ and A. Lef\`evre$^3$}
\affiliation{
(1) DAMTP, CMS, University of Cambridge, Cambridge, CB3 0WA, UK \\
(2) IRSAMC, Laboratoire de Physique Th\'eorique, Universit\'e Paul Sabatier, 
118 route de Narbonne, 31062 Toulouse Cedex 04, France\\
(3) Department of Theoretical Physics, 1 Keble Road, Oxford, OX1
  3NP, UK}

\begin{abstract}
The behavior of the self diffusion constant of Langevin 
particles interacting via a pairwise
interaction is considered. The diffusion constant is calculated
approximately within a perturbation theory in the potential strength
about the bare diffusion constant. It is shown how this expansion leads
to a systematic double expansion in the inverse temperature $\beta$
and the particle density $\rho$. The one-loop diagrams in this expansion
can be summed exactly and we show that this result is exact in the 
limit of small  $\beta$ and $\rho\beta$ constant. The one-loop result
can also be re-summed using a semi-phenomenological renormalization group 
method which has proved useful in the study of diffusion in random media.
In certain cases the renormalization group calculation predicts the 
existence of a diverging relaxation time signalled by the vanishing of the
diffusion constant -- possible forms of divergence coming from this
approximation are discussed.  Finally, at a more quantitative level, 
the results are compared with numerical simulations, in two-dimensions,
of particles interacting via a soft potential recently used to model 
the interaction between coiled polymers. 
\end{abstract}
\begin{center} 
\date{24 February 2003}
\end{center}

\maketitle


\newcommand{\bq}{\ensuremath{{\bf q}}}
\newcommand{\bbq}{\ensuremath{{\bf Q}}}
\newcommand{\bp}{\ensuremath{{\bf p}}}
\newcommand{\bk}{\ensuremath{{\bf k}}}
\newcommand{\br}{\ensuremath{{\bf r}}}
\newcommand{\bx}{\ensuremath{{\bf x}}}
\section{Introduction}
Transport properties of deeply cooled liquids may change by several
orders of magnitude when one reduces the temperature
\cite{angel}. Indeed, a liquid
undergoing a deep quench below its melting transition may stay in a
metastable supercooled state. In this state, the relaxation time is
much larger than the experimental time scale, and the system
is always out of equilibrium. Experiments suggest that in 
``fragile'' glass formers, there is a temperature $T_0$ at which the
relaxation time diverges. Even if the mere existence of a divergence
of the relaxation time $\tau$ at finite temperature is still controversial, the
experimental data can be often fitted by the so-called
Vogel-Fulcher-Tammann law $\tau\sim \exp\left(\frac{A}{T-T_0} \right)$.
Since such a glassy behavior has been observed in a wide range of
materials, there have been huge efforts dedicated to computing the  transport
properties in supercooled liquids. Among them, the Mode Coupling
Theory (MCT) \cite{ben,leuth,gotze} appears to give results and
predictions which fit remarkably well with 
data for many different systems \cite{kob}. However, despite 
its successes, the derivation of MCT either from the Mori-Zwanzig
formalism \cite{gotze} or from fluctuating hydrodynamics \cite{das} 
remains quite obscure, and systematic improvements seem difficult
to implement and control.    
Hence, with the aim of understanding better the structure of the
dynamics, alternative methods of computing transport properties of systems
of many interacting particles such as supercooled liquids 
are worth examining. 
In this paper, we will develop a method to compute
the long time self-diffusion constant which allows systematic
double perturbative expansions, in the strength of the interaction and
in the density of particles. The approach is based on the Langevin
dynamics for $N$ particles with two-body interactions, though
this approach may be generalised easily to three-body
interactions. Such Markovian Langevin dynamics can be invoked in
liquids over length and time scales where inertial effects become
negligible. The Langevin equation thus represents a course grained
image of the system and the effective parameters and interactions
used in the Langevin approach require microscopic derivation. 
The Langevin approach also naturally describes the dynamics of
colloids in solution \cite{rusav}, the Brownian noise is induced by the
solvent and the effective interaction between particles is composed
of a pairwise direct interaction between the colloids plus additional
hydrodynamic interactions induced via the solvent.

We consider the interacting set of Langevin equations for particles
${\bf X}_i$ interacting via a pairwise potential 
$V({\bf X}_i- {\bf X}_j)$ depending only on
the distance between the particles at 
temperature $T$ in $D$ dimensions:

\begin{equation}
\frac{d{\bf X}_i^\alpha}{dt} = 
-\lambda\sum_{j} \partial_{X^\alpha_i} V({\bf X}_i - {\bf X}_j) + 
\eta_i^\alpha.
\end{equation}
The units of time are chosen such that the white noise field 
$\eta_i^\alpha$ has correlation function
\begin{equation}
\langle \eta_i^\alpha(t) \ \eta_j^\beta(t')\rangle = 2\kappa \delta_{ij}
\delta^{\alpha \beta}\delta(t-t'),
\end{equation}
where the angled brackets indicate averaging over the thermal noise.
The term $\kappa$ is thus the bare diffusion constant of the 
particles in the absence of interactions. The fluctuation dissipation
theorem or Einstein relation implies that ${\lambda/\kappa} = 1/T$.
This system of equations can be used to describe a colloidal system
of interacting particles suspended in solution when hydrodynamic interactions
are neglected. The neglecting of hydrodynamic interactions
is justified where the direct two body interaction $V$ is of much longer range
than the hydrodynamic interactions. A commonly cited example
is charged colloids in a solution when the Debye length is very large
with respect to the particle sizes.

The effective macroscopic diffusion constant 
$\kappa_e$ of particle $i$ is defined by
\begin{equation}
\lim_{t\to \infty}\langle X_i^2(t) \rangle = 2 D \kappa_e t.
\label{defk}
\end{equation}

It is the aim of this paper to develop a new technique for the calculation
of $\kappa_e$. There are two basic routes to calculate $\kappa_e$.
The first is based on the  direct calculation of the diffusion constant
\cite{hekl,acme,yeme,yeac,acfl1,fejo,haheke,medina,nabedh,ac,made,inra}.
The second is based on the
calculation of the modification of the response to a small external force
on a given particle due  the interaction with the other particles - 
the so called relaxation effect; the resulting value of
$\kappa_e$ is then determined from the Einstein or fluctuation 
dissipation relation \cite{bat,szle,lesz,ledh}.
One approach to studying the dynamics of a tracer particle is to write
an effective one particle  Langevin equation with a non-Markovian 
memory kernel, derived via projection operator techniques
\cite{hekl,acfl1,ac,ledh,inra,nabedh,acme,fejo,yeme,yeac}. 
This kernel must then be computed by invoking approximation or closure 
schemes such as  Mode Coupling-like approximations \cite{medina,nabedh},
cluster expansions \cite{fejo}, or weak coupling expansions 
\cite{hekl,ac,made}. Other approaches are based on
closure schemes for the Smoluchowski equation 
\cite{szle,acme,lesz,yeme,yeac}, which normally
involve closing the hierarchy of equations 
for the joint probability density functions
by replacing, for instance, the three body joint probability density function
by its corresponding Kirkwood superposition approximation. 
Both techniques can be handled to produce results which are 
exact to first order in the particle
density $\rho$ \cite{bat,acfl1,ledh,fejo,haheke}. 

\section{Diagrammatic expansion}
\begin{figure*}
\includegraphics[width=.7\textwidth]{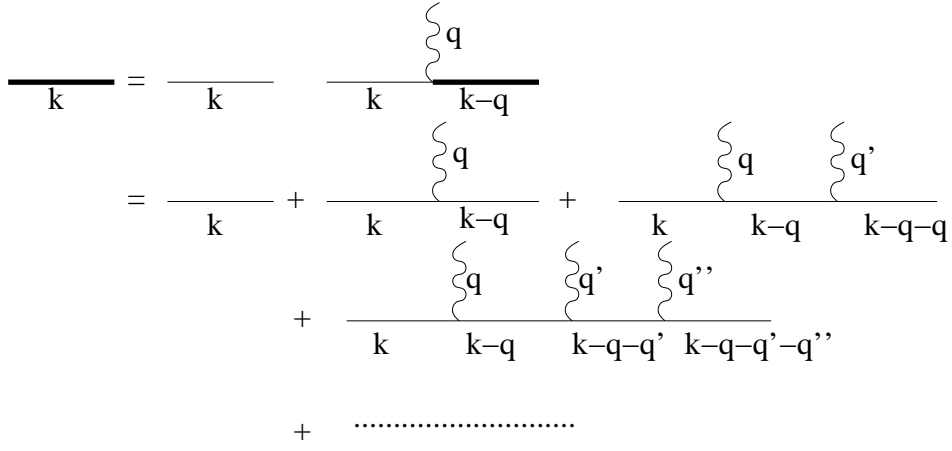}
\caption{Diagrammatic expansion for ${\tilde P}(\bk,s)$}
\label{fig1}
\end{figure*}

Here we use a technique based on a perturbative weak coupling 
expansion of the Smoluchowski or Fokker-Planck equation for the $N$ particles
in interaction. The form of this perturbation expansion is
identical  to that used to calculate the effective diffusion
constant of a particle in a random potential. We denote by
$P({\bf x}_1,{\bf x}_2 \cdots, {\bf x}_N,t)$ the probability density function
for the particle displacements from their original positions at $t=0$ at time 
$t$, that is to say the density of $\left\{ {\bf X}_i(t)-{\bf X}_i(0);\ \ \
1\leq i\leq N\right\}$. We take initial conditions 
$ {\bf X}_i(0) = {\bf x}^{(0)}_i$
where the ${\bf x}^{(0)}_i$ are independently and uniformly 
distributed throughout the volume $V$ of the system. In principal 
one could take other initial conditions for the  ${\bf x}^{(0)}_i$, notably
one could take their equilibrium distribution. However if the system is
ergodic then the resulting behavior of the diffusion effective constant 
should be independent of this distribution. The disorder induced by the 
random initial conditions shows the link with diffusion in a random
potential.   
The forward Fokker Planck or generalized Smoluchowski  equation for $P$ is   
\begin{equation}
\frac{\partial P}{\partial t}
= \kappa \nabla^2 P + \lambda \nabla\cdot(P\nabla \phi),
\label{eqfp}
\end{equation}
where $\nabla$ is the gradient operator on $\mathbb{R}^{DN}$
where $D$ is the spatial dimension and $N$ the number of particles.
The potential $\phi$ in this formalism is given by 
\begin{equation}
\phi({\bf x}_1,{\bf x}_2 \cdots, {\bf x}_N) = \phi_0({\bf x}_1+{\bf x}_1^{(0)} ,
{\bf x}_2 + {\bf x}_2^{(0)}\cdots, {\bf x}_N+{\bf x}_N^{(0)} ),
\end{equation}
where
\begin{equation}
\phi_0({\bf x}_1,{\bf x}_2 \cdots, {\bf x}_N) =
\sum_{i<j} V({\bf x}_i -{\bf x}_j).
\end{equation}
From here on we shall denote by the vector ${\bf x}$, without a particle index,
the global position vector $({\bf x}_1, {\bf x}_2, \cdots {\bf x}_N)$ in 
$\mathbb{R}^{DN}$ and by the vector $\bk$ the corresponding Fourier 
vector $(\bk_1, \bk_2, \cdots \bk_N)$. If one defines
\begin{equation}
{\tilde P}(\bk,s) = \int_0^\infty dt\int_{\mathbb{R}^{DN}}
d{\bf x} \exp(-st-i\bk\cdot{\bf x}) P({\bf x},t),
\end{equation}
it is straightforward to show that ${\tilde P}(\bk,s)$
obeys
\begin{equation}
\begin{split}
&{\tilde P}(\bk,s) = \frac{1}{\kappa \bk^2 + s}\\
&- \frac{\lambda}{\kappa \bk^2 + s}\int_{\mathbb{R}^{DN}}
\frac{d\bq}{(2\pi)^{ND}}\ \bk\cdot\bq {\tilde\phi}(\bq)
{\tilde P}(\bk-\bq,s).
\end{split}
\end{equation}
We note that because of our choice of coordinates relative to the initial 
conditions, $P({\bf x},0) = \delta({\bf x})$ and also
\begin{equation}
{\tilde\phi}(\bq) = \exp(i\bq\cdot{\bf x}^{(0)}){\tilde\phi}_0(\bq).
\end{equation}
One has also that
\begin{equation}
{\tilde\phi}_0(\bq) = (2\pi)^{(N-1)D}{\tilde\psi}_0(\bq),
\end{equation}
with 
\begin{equation}
{\tilde\psi}_0(\bq)
= \sum_{i< j} {\tilde V}(\bq_i)\delta(\bq_i + \bq_j)
\prod_{k\notin \{i,j\}}\delta(\bq_k).
\label{eqpsi0}
\end{equation}
Using these definitions one obtains the equation
\begin{widetext}
\begin{equation}
{\tilde P}(\bk,s) = {1\over \kappa \bk^2 + s}
- {\lambda\over \kappa \bk^2 + s}\int_{\mathbb{R}^{DN}}
{d\bq\over (2\pi)^{D}}\ \bk\cdot\bq {\tilde\psi}_0(\bq)
\exp(i\bq\cdot{\bf x}^{(0)})
{\tilde P}(\bk-\bq,s).
\label{fpft}
\end{equation}
\end{widetext}
Eq. (\ref{fpft}) can be solved iteratively leading to the expansion
which is represented in Fig. (\ref{fig1}). We note here that this weak
coupling expansion as it stands only makes sense for potentials which are 
bounded as $V({\bf x})$ 
is treated as a small perturbation. However the series can
be re-summed to obtain physical results for unbounded (such as hard core)
potentials. In addition the behavior of soft potentials is of direct
physical interest as they provide course grained descriptions of
coiled polymers \cite{cp}, star polymers \cite{sp}
and  micelles \cite{mic} in solvents.

Momentum is conserved at
each vertex and the Feynman rules are
\begin{itemize}
\item Each solid line (horizontal) corresponds to the bare propagator
\begin{equation}
G_0(\bk,s) = {1\over \kappa \bk^2 + s},
\end{equation}
where $\bk$ is the momentum in that line.
\item Each vertex (vertical wavy line) carries a factor
\begin{equation}
-\lambda \bk\cdot\bq\ \psi_0(\bq) \exp(i\bq\cdot{\bf x}^{(0)}),
\end{equation}
where $\bk$ is the in going momentum (from the left) and $\bq$ 
is the momentum flowing into the vertical wavy line.
\item Each  momenta $\bq$ flowing into a wavy line is integrated
over ${\mathbb{R}^{DN}}$ with the measure $d\bq/(2\pi)^D$.
\end{itemize}

We see that the ingoing and outgoing momentum on each of the generated
diagrams is not conserved. This is because the spatial translational 
invariance of the system is not explicit. To study a spatially translational
invariant system we average over the initial position ${\bf x}^{(0)}$
throughout the volume $V^{N}$. We shall take uniform initial conditions 
although any initial conditions which have the property of spatial 
translational invariance should give the same asymptotic 
(late time) properties 
for $G(\bk,s) = \langle {\tilde P}(\bk,s) \rangle_0$. Here the
angled brackets with the $0$ subscript indicate the average over the initial
position vector ${\bf x}^{(0)}$ and is defined by
\begin{equation}
\langle A  \rangle_0 = {1\over V^N}\int_{V^N} d{\bf x}^{(0)} A({\bf x}^{(0)}).
\end{equation}
In the limit of large $V$ this integration multiplies each
diagram by a factor of $(2\pi)^{ND}\delta(\sum_v \bq_v)/V^N$
where the $\bq_v$ are all the momenta flowing into the upward
wavy lines in each diagram at each vertex $v$. 
This momentum conservation ensures that
the momentum flowing into each diagram is the same as that flowing out and
indicates the invariance by translation in space of the system averaged
over its initial conditions.
After taking this average over the diagrams in Fig. (\ref{fig1})
we obtain the diagrammatic expansion for $G(\bk,s)$. The Feynman rules
are as before but with the following modifications
\begin{itemize}
\item Each diagram carries an overall 
factor of $(2\pi)^{(N-1)D}/V^N$. 
\item There are only $n-1$ independent
momenta for each diagram of $n$ vertices by momentum conservation. 
Each of these momenta is integrated with the measure  $d\bq/(2\pi)^D$
as before.
\end{itemize}
\begin{figure*}
\includegraphics[width=.7\textwidth]{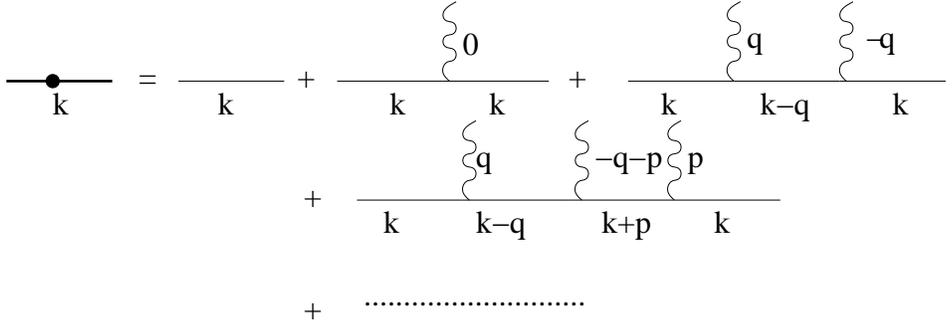}
\caption{Diagrammatic expansion for $G(\bk,s)$ 
(shown as a line with a blob) 
obtained after averaging over initial particle positions in the diagrammatic
expansion shown in Fig. (\ref{fig1}).}
\label{fig2}
\end{figure*}

The form of the perturbation expansion shown in Fig. (\ref{fig2}), shows that
one can write $G(\bk,s)$
\begin{equation}
G(\bk, s) = {1\over \kappa \bk^2 + s} +{1\over (
\kappa \bk^2 + s)^2} D(\bk,s)
\end{equation}
The term $D(\bk,s)$ can clearly be expressed in terms of 
one-particle-irreducible diagrams. Diagrams which are one-particle-reducible 
are those containing a propagator having a pure momentum $\bk$ flowing through
one of their propagators and can thus be factorised. 
The propagator connecting two one-particle-irreductible diagrams is therefore 
the bare one $1/(\kappa \bk^2 +s)$.
If we denote the one-particle-irreducible diagram contribution as
$\Sigma(\bk,s)$ then
\begin{equation}
G(\bk, s) = {1\over \kappa \bk^2 + s} + {\Sigma(\bk,s)\over 
(\kappa \bk^2 + s)^2} + {\Sigma(\bk,s)^2\over 
(\kappa \bk^2 + s)^3} + \cdots
\end{equation}
which sums to give
\begin{equation}
G(\bk, s)  = {1\over \kappa \bk^2 -\Sigma(\bk,s) + s}
\end{equation}
We now note that by the conservation of probability $G(0, s)=1/s$ and we
thus expect that for small $|\bf k|$
\begin{equation}
G(\bk, s) = {1\over \kappa \bk^2 - \bk^2 E(s) + s}
\end{equation}
where $\Sigma(\bk,s) \approx \bk^2 E(s)$. In the limit of small $|\bk|$
and $s$ therefore
\begin{equation}
G(\bk, s) = {1\over (\kappa- E(0)) \bk^2   + s}
\label{eqgke}
\end{equation}

The effective diffusion constant for the particles is extracted 
using the fact that
\begin{equation}
2D\kappa_e t = \lim_{t\to \infty}\langle({\bf X}_i(t)
- {\bf x_i}^{(0)})^2\rangle,
\end{equation}
where the angled brackets on the right hand side above indicate the average
over the thermal noise and over the initial conditions.
We also have that
\begin{equation}
G(\bk,s) = \langle \exp\left(
-i\sum_i\bk_i \cdot({\bf X}_i(t) - {\bf x}_i^{(0)})\right)\rangle
\label{eqfte}
\end{equation} 
For small $s$ in Laplace space we have 
\begin{eqnarray}
\int_0^\infty &dt&\ \exp(-st) \langle ({\bf X}_i(t) - {\bf x}_i^{(0)})^2\rangle
\nonumber \\ 
&\approx&\int_0^\infty dt\ \exp(-st)2D\kappa_e t  
= {{2 D \kappa_e}\over s^2} 
\end{eqnarray}
and therefore, from Eq. (\ref{eqfte}), for small $s$
\begin{equation}
{{2 D \kappa_e}\over s^2} \approx -\sum_{\alpha = 1}^D {\partial^2
\over \partial {k_i^\alpha}^2}G(\bk,s)\vert_{\bk=0}.
\end{equation}
Now using Eq. (\ref{eqgke}) we obtain
\begin{equation}
\kappa_e = \kappa - E(0)
\end{equation}
The calculation of $\kappa_e$ can therefore be evaluated from $G(\bk,0)$. 
We may express the term $E(0)$ as an expansion in the number of vertices
$\nu$ in the diagram, we write
\begin{equation}
E(0) = \sum_\nu E_{\nu}
\end{equation}
where 
\begin{equation}
E_{\nu} = \lim_{|\bk|\to 0}  {1\over  \bk^2}\Sigma_\nu(\bk,0)
\end{equation}
where $\Sigma_\nu(\bk)$ is the sum of one-particle-irreducible diagrams
with $\nu$ vertices.
From Fig. (\ref{fig2}) we may write
\begin{equation}
G(\bk,0) = {1\over \kappa \bk^2} +{1\over (\kappa \bk^2)^2}
\sum_\nu D_\nu(\bk,0)
\end{equation}
where $D_\nu(\bk,0)$ is the sum over all terms with $\nu$ vertices and
not just the one-particle-irreducible ones. The one-particle-irreducible
components of $D_\nu(\bk,0)$, $\Sigma_\nu(\bk,0)$, must be extracted 
from $D_\nu(\bk,0)$. The terms $D_\nu(\bk,0)$ have the behavior
$D_\nu(\bk,0) \approx F_\nu \bk^2$ for small $|\bk|$.
It is straightforward to verify that the term $F_1$ is zero. This means
therefore that there is no contribution to $O(\lambda)$ in the 
asymptotic single particle diffusion constant $\kappa_e$. 
Examining  Fig. (\ref{fig2}) we find that 
\begin{equation}
\begin{split}
D_2(\bk,0) &= {(2\pi)^{D(N-2)}\over \kappa V^N} 
\lambda^2\\
&\int_{\mathbb{R}^{DN}}d\bq\ 
{\bk\cdot\bq\ (\bk - \bq) 
\over (\bk - \bq)^2 }\psi_0(\bq)\psi_0(-\bq).
\label{eqd2}
\end{split}
\end{equation}

\begin{figure}
\includegraphics[width=.45\textwidth]{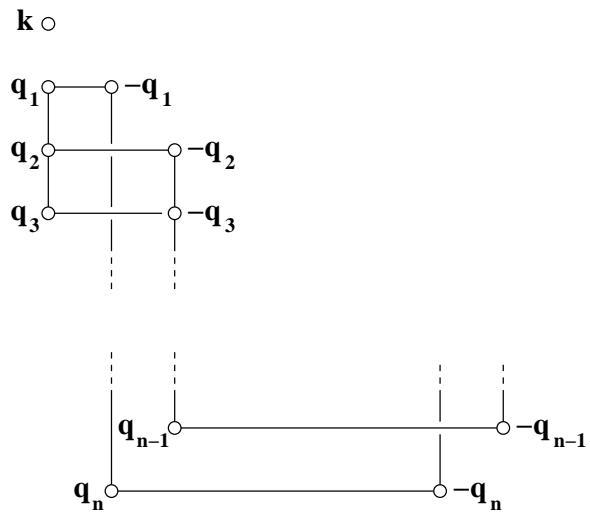}
\caption{Diagrammatic representation of the vertex momentum present
in one term of the pair development of a product of $\psi_0$'s. }
\label{fig3}
\end{figure}

The potential term in the integrand may be expressed using Eq. (\ref{eqpsi0})
as
\begin{equation}
\begin{split}
&\psi_0(\bq)\psi_0(-\bq) = \sum_{i< j}\sum_{k< l}
{\tilde V}(\bq_i){\tilde V}(-\bq_k)
\\ \delta(\bq_i + \bq_j)\delta(\bq_k + \bq_l)
&\times\prod_{r\notin \{i,j\}} \delta(\bq_r)
\prod_{s\notin \{k,l\}} \delta(\bq_s)
\label{eqnpotpot}
\end{split}
\end{equation}
One must keep in mind that only terms with nonzero
$\bq$ can contribute (one can imagine an additional  
term $+s$ in the denominator
for small $s$ which is taken to zero at the end of the calculation).
The only terms in Eq. (\ref{eqnpotpot}) which have non-zero momentum are those 
where the pair $(i,j) = (k,l)$. 
In addition, the computation is simplified if one assumes $\bk_r=0$
for $r>1$. Clearly the coefficient of $\bk_1^2$ in $\kappa \bk^2 -\Sigma
(\bk,0)$ is the self diffusion constant of particle $1$ which is also the self
diffusion constant of any given particle. Hence for the purposes
of the calculation of $\kappa_e$ we can restrict our selves to the case
$\bk = (\bk_1,0,0\cdots 0)$.
In this case, the only  choices of the particle indices giving 
a nonzero diagram are  $i=k=1$, and $j=l$ due to the scalar product 
$\bk \cdot \bq$ on the first vertex.
There are thus $N-1$ identical non-zero diagrams with two vertices. 
For notational simplicity in the following we will  write $\bk=\bk_1\in
\mathbb{R}^D$. All the choices are equivalent to choosing
$\bq_1=\bq$, $\bq_2=-\bq$ and  $\bq_r=0$ for $r>2$ where again 
$\bq \in \mathbb{R}^D$.
In the resulting integral there are repeated delta functions for the terms 
$r=s$ and also a double $\delta(\bq_i + \bq_j)$. 
We use the relation for $\bq \in \mathbb{R}^{D}$
\begin{equation}
\delta^2(\bq) = \delta(\bq) {V\over (2\pi)^D}.
\end{equation}
The result for $D_2(\bk)$ is thus
\begin{equation}
D_2(\bk,0) = {\lambda^2\rho\over
\kappa}\int_{\mathbb{R}^{D}}{d\bq\over (2\pi)^D}\ \frac{{\bf 
k}\cdot \bq\ (2\bq-\bk)\cdot \bq}{(\bk-\bq)^2+{\bf
q}^2}\ {\tilde V}(\bq)^2,
\end{equation}
The above diagram is also clearly one-particle-irreducible and hence 
$F_2 = E_2$ thus giving
\begin{equation}
E_2 = {1\over 2}{\lambda^2\rho\over \kappa D}\int_{\mathbb{R}^{D}}
{d\bq\over (2\pi)^D}\ {\tilde V}(\bq)^2 
= {1\over 2}{\lambda^2\rho\over \kappa D}\int_{\mathbb{R}^{D}}
d{\bf x}\ V^2({\bf x}).
\label{eqnE2}
\end{equation}
The diagram giving $D_3(\bk,0)$  has the value
\begin{widetext}
\begin{equation}
D_3(\bk,0) = -{(2\pi)^{D(N-3)}\over \kappa^2 V^N} \lambda^3
\int_{\mathbb{R}^{DN}} d\bq\   d\bp\ 
{\bk\cdot\bq\ (\bk - \bq)\cdot (\bq + \bp)
\ (\bk + \bp)\cdot\bp \over (\bk -\bq)^2
(\bk + \bp)^2} \psi_0(\bq)\psi_0(\bp)\psi_0(-\bq- \bp).
\end{equation}
\end{widetext}
To simplify the counting of diagrams with nonzero momentum let us 
consider the general expansion of $\psi_0({\bf q_1})\cdots
\psi_0({\bf q_n})$ in a diagram with $n$ vertices. From Eq. (\ref{eqpsi0})
a given term on expanding the $n$-fold sum over pairs has the form
\begin{equation}
A_{i_1 j_1}(\bq_1)\cdots A_{i_n j_n}(\bq_n).
\end{equation}
Any diagram where the momentum $\bq_v$ flowing into the vertex $v$
is zero due to the presence of the scalar product with  $\bq_v$
at each vertex in the Feynman rules. The momentum flowing into the vertex
$v$ in the above decomposition over pairs is $(0,0,\cdots \bq_i
\cdots -\bq_i \cdots 0,0)$ {\em i.e.} it has $\bq_i$ at
the particle position $i$ and $-\bq_i$ at the particle position $j$
in the total momentum vector. This is shown diagrammatically
in Fig. (\ref{fig3}). 

\begin{figure}
\includegraphics[width=.14\textwidth]{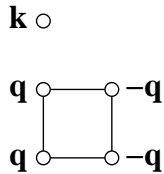}
\caption{Diagram in the pair development contributing to $E_2$}
\label{fig4}
\end{figure}

\begin{figure*}
\includegraphics[width=.2\textwidth]{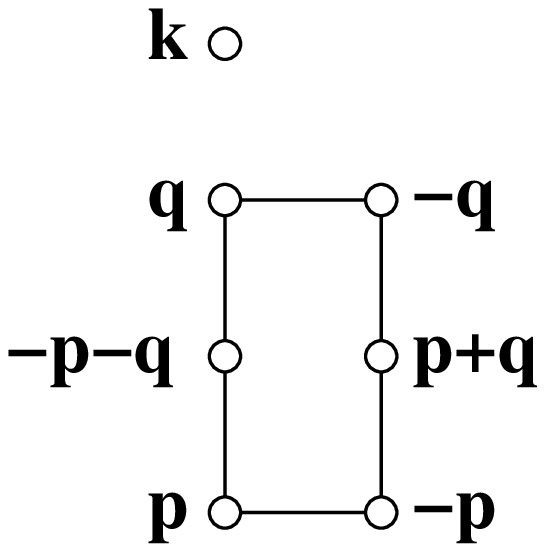}
\includegraphics[width=.2\textwidth]{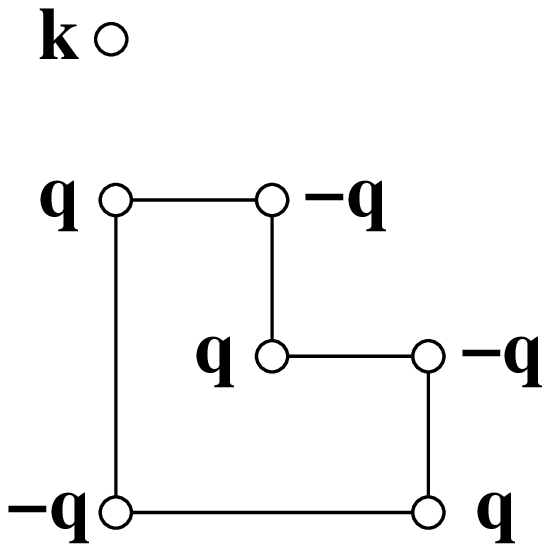}
\includegraphics[width=.2\textwidth]{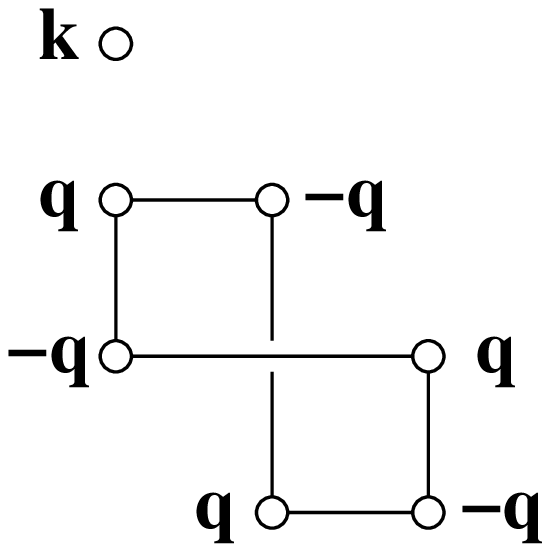}
\caption{Diagrams in the pair development contributing to $E_3$ (left: type
one, center: type 2, right: type 3)}
\label{fig5}
\end{figure*}

Each horizontal line corresponds to a vertex and the horizontal coordinates
are given by the points $(i_v,j_v)$. Each line must have a non zero momentum,
thus $\bq_v \neq 0$. However the sum of the momenta down each column
must also be  zero. Hence each coordinate $i_v$ must appear on at least
two lines in order to give a diagram which is nonzero. In 
Fig. (\ref{fig4}) one example of the equivalent nonzero diagram 
contribution to $E_2$ is shown.
The 3 types of diagrams 
which give nonzero contributions to $F_3$ are shown in Fig. (\ref{fig5}). 
The diagrams
of type 1 have a multiplicity of $N-1$ and have the same particle pair on 
each line. For the diagrams
of type 2 and 3 there are $(N-1)(N-2)$ diagrams of the precise form shown in 
the diagram. This is because there are $N-1$ choices for the first pair $(1,2)$
and $N-2$ choices of the pair $(2,3)$ (that is to say the particle
number $3$) shown on the second line. The choice of the 
last (third) pair is not free  by momentum conservation. 
Due to the presence of a delta function $\delta(\sum_v \bq_i^v)$
for each column which is automatically satisfied when all the $\bq_i^v$
are zero, each empty column carries a factor of $V/(2\pi)^D$. Hence
a diagram with $k$ empty columns has a factor
$\left(\frac{V}{(2\pi)^D}\right)^k$. In diagrams of type 1, $k= N-2$ and in
diagrams of type 2 and 3, $k = N-3$.  

The contribution of diagrams of type 1 to $D_3(\bk,0)$ is thus
\begin{widetext}
\begin{equation} 
D_3^{(1)}(\bk,0) = -{\lambda^3\rho\over \kappa^2} \int_{\mathbb{R}^{2D}}
{d\bq\over (2\pi)^D}{d\bp\over (2\pi)^D}\
\frac{\bk\cdot \bq\ [(2\bq-\bk)\cdot (\bp+\bq)]\ [(\bk +2\bp)\cdot
\bp]}{[(\bk-\bq)^2 +\bq^2]\ [(\bk+\bp))^2+\bp^2]}\
{\tilde V}(\bq){\tilde V}(\bp){\tilde V}(\bp+\bq), 
\end{equation}
which gives
\begin{equation} 
F_3^{(1)} = -{\lambda^3\rho\over 2 D \kappa^2} \int_{\mathbb{R}^{2D}}
{{d\bq\over (2\pi)^D}{d\bp\over (2\pi)^D}
{\tilde V}(\bq){\tilde V}(\bp){\tilde V}(\bp+\bq) 
(1 - {({\bf p\cdot  q})^2\over \bp^2 \bq^2})}.
\end{equation}
\end{widetext}
The contribution from diagrams of type 2 and 3 is
\begin{equation} 
F_3^{(2)} = -{3\lambda^3\rho^2\over 4D \kappa^2} \int_{\mathbb{R}^{D}}
{d\bq\over (2\pi)^D}
{\tilde V}(\bq)^3
\end{equation}
These diagrams contributing to $F_3$ are also one-particle-irreducible 
and hence 
to $O(\lambda^3)$, we obtain
\begin{widetext}
\begin{equation}
\frac{\kappa_e}{\kappa}=1 - 
{\rho \lambda^2 \over 2D \kappa^2}\int_{\mathbb{R}^{D}} 
{d\bq\over (2\pi)^D} {\tilde V}(\bq)^2   
 + {\rho \lambda^3\over 2D \kappa^3}
\int_{\mathbb{R}^{2D}}{{d\bq\over (2\pi)^D}{d\bp\over (2\pi)^D}
{\tilde V}(\bq){\tilde V}(\bp){\tilde V}(\bq+\bp) 
(1 - {({\bf p\cdot  q})^2\over \bp^2 \bq^2})}+ 
{3\rho^2 \lambda^3\over 4D\kappa^3} \int_{\mathbb{R}^{D}}
{d\bq\over (2\pi)^D}
{\tilde V}(\bq)^3 
\label{eqkappa1}
\end{equation}
\end{widetext}
Hence the ratio $\kappa_e/\kappa$ is expressed as a perturbation expansion in 
$\frac{1}{T}$. Notice here that there are more 
lines than rows in the diagrams 
of the pair development, which means that a finite number of diagrams
contribute to any given order of the $\frac{1}{T}$-expansion, 
whereas an infinite number of diagrams must be summed in order to compute 
any order in the $\rho$-expansion. A diagram with $n$ lines and $m$ 
non-empty columns is of order $\rho^{m-1}(\lambda/\kappa)^n
= \rho^{m-1}\beta^n$, thus a systematic double expansion in $\rho$ and
$\beta$ may be performed.  We note that in Eq. (\ref{eqkappa1})
the leading order term in $\beta$ which is of order $\rho\beta^2$ 
recovers the weak coupling approximation. It is clear that the 
weak coupling approximation is not valid at low densities if
the temperature is too low. The calculation of $\kappa_e$ to first
order in $\rho$ involves summing all the diagrams which have 
just two columns occupied. These diagrams, which are all 
one-particle-irreducible, can be re-summed 
via an  integral equation \cite{deleip} and one can 
show that the resulting expression for $\kappa_e$ is the same as 
that given by the relaxation method applied to the effective two-body 
problem as expounded in \cite{ledh}. 

\begin{figure}[htbp]
\includegraphics[width=.28\textwidth]{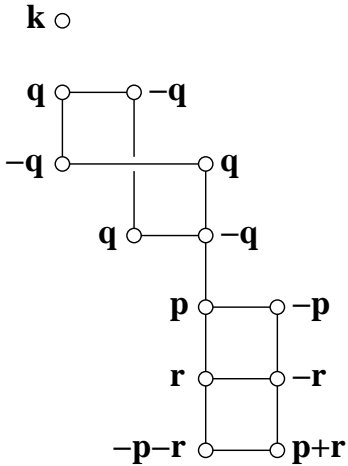}
\caption{Example of a diagram with disconnected loops in the pair
development. Here the $\bq$-loop is disconnected from the $(\bp,\br)$-loop.}
\label{fig6}
\end{figure}

Consider a diagram with $v$ vertices and its $\psi_0$ pair expansion.
If the first $v'$ vertices do not contain more than one particle index 
in common
with the $v-v'$ remaining vertices then the momentum flowing between 
the vertex $v$ and $v+1$ is zero (or {\bk} if the column
in common is the first one) and thus the  diagram is zero (or one particle 
reducible). In other words,
all diagrams with disconnected loops such as the one shown 
in Fig. (\ref{fig6}) have a zero value or are one-particle-reducible and thus
give no contribution to $\kappa_e$.

As an example, we consider potentials of the form
\begin{equation}
V({\bf r}) = {\epsilon\over (2\pi)^{D\over2}} \exp\left(-{{\bf r}^2\over 
2r_0^2}\right),
\label{potgauss}
\end{equation}
which has been recently proposed to model the effective interaction between 
coiled polymers \cite{cp} at weak dilution. 
The Fourier transform of $V({\bf r})$ is then given by
\begin{equation}
 {\tilde V}(\bq) = {\epsilon r_0^D}
\exp\left(-{\bq^2 r_0^2\over 2}\right).
\end{equation}
With this interaction potential and setting $\epsilon = r_0 =1$ we
find from Eq. (\ref{eqkappa1})
  
\begin{equation}
F_q^{(p)}= a_{pq} \rho^pT^{-q},
\label{eqkappas}
\end{equation}
where the first non zero $a_{pq}$'s are
\begin{equation}
\begin{split}
a_{12}&= -\frac{1}{2^{D+1}D\pi^{\frac{D}{2}}},\\
a_{13}&=\frac{1}{2D}\left(\frac{1}{3^\frac{D}{2}}-\frac{K(D)+D K(D+2)}{(2
\pi)^D} \right), \\
a_{23}&= \frac{3}{4 D (6\pi)^{D/2}},\\
K(D)&=\int_0^{\frac{1}{\sqrt{3}}}\,dx\,\frac{x^{D-1}}{1+x^2}.
\label{eqkappas1}
\end{split}
\end{equation}
Clearly this $\frac{1}{T}$-expansion is valid only at very high
temperature. It can be improved in several manners. One could 
compute higher orders in this expansion, 
which should give better agreement 
with the simulations, but however the expansion will inevitably break down  
at low $T$. Alternatively, 
one could  try to sum infinite sub-series in order to build approximate 
non-perturbative schemes. As mentioned above one approach is to try and re-sum
the diagrams to obtain results exact for all $\beta$ to order  $\rho$. Here
we shall concentrate on another re-summation involving only one-loop diagrams. 

\section{One-loop analysis}

\subsection{Simple one-loop contribution}
Here we will focus on the class of one-loop diagrams. These diagrams
are those which involve only one momentum integral. Is this case, there are
two dots on each occupied line or row and these diagrams are 
all one particle-irreducible. In addition from the discussion
in the previous section, these diagrams are the dominant ones
in the limit where $\rho\beta = c $ (with $c$ a constant)
and  $\rho \to \infty$,
or equivalently where $\rho\beta = c $ 
and $\beta \to 0$. The dimensionless form of $c$ would in fact be
$c'= \rho r_0^D \beta\epsilon $, where $r_0$ is the characteristic
length scale of the potential and $\epsilon$ its energy. 
As before, we write $\bk=\bk_1\in\mathbb{R}^D$. The interesting point about 
this limit is that the (high temperature) statics of the model can also
be evaluated \cite{acsa}, only chain diagrams in the virial expansion 
are retained in this limit. In electrolytic systems the  
Debye-H\"uckel approximation is recovered on the retention of only chain 
diagrams \cite{hamc} and hence it is interesting that one can have a theory of 
$\kappa_e$ for electrolyte systems which is compatible with the 
Debye-H\"uckel approximation which has proved so useful in the study of their
static properties.

\begin{figure*}
\includegraphics[width=.6\textwidth]{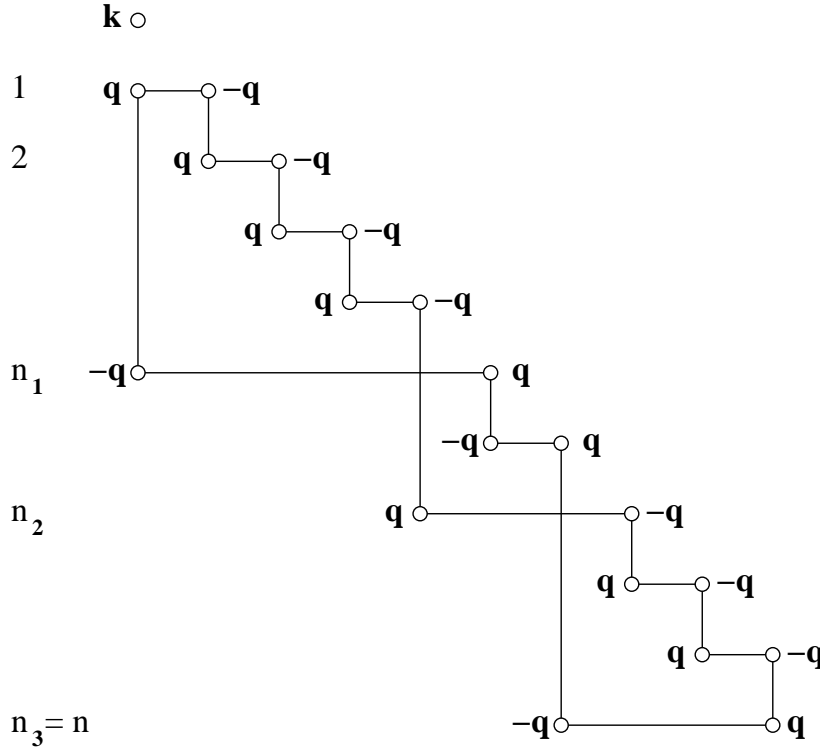}
\caption{Example of a staircase diagram contributing to the one-loop 
expansion.}
\label{fig7}
\end{figure*}

The typical one-loop diagrams
can be reduced to the staircase like diagrams shown in Fig. (\ref{fig7}) by
relabelling the particles. We note here that as we are summing an 
infinite number of diagrams one must be careful not to include the
same particle twice in the same diagram as this will make the 
diagram two-loop. Hence for a finite system one cannot have a
one loop diagram with more than  $N$ vertices as it will 
include at least one particle (other that the tracer particle (1)) at 
least three times, meaning that the diagram can have at least two
two independent momenta flowing through it. In this
case the counting of the one-loop diagrams with $n$ vertices where
$n$ is of order $N$ will be different. We have however taken the limit
$N\to \infty$ already and hence this does not cause us any problems in the 
region where the power series giving the one-loop result is convergent. 

In constructing a one-loop diagram at any level one must insert 
two points one of which must coincide with one of the two unpaired
(in the vertical direction) points above. The leftmost point
of the new pair can either be paired with the rightmost of the 
two unpaired points above or it can be paired with the leftmost 
of the two unpaired points above - this gives a crossing. Clearly 
the first crossing must be back to the first  column. 
Each diagram thus has $p$ crossings where $p\in [1,n-1]$. If there
is only one crossing then it must occur at level $n$ corresponding
to a  staircase diagram with one crossing at the last line. 
Up to particle relabelling there is 
only one such diagram and its contribution is
\begin{equation}
b_1(n)=\frac{2n-3}{2^{n-1}D}\frac{\rho^{n-1}(-\lambda)^n}{\kappa^{n-1}}
\int_{\mathbb{R}^{D}} {d\bq\over (2\pi)^D}\ {\tilde V}(\bq)^n,
\end{equation}
If $n>2$ and $p>1$ then the first crossing must occur at the level 
$n_1 \in [2,n-1]$. At the level $n_1 +1$ the next two points can either
cross or not cross just till the level $n$ where the diagram most close.
There are thus $2^{n-1-n_1}$ topologically distinct diagrams with the 
same first crossing point at $n_1$. 
It is easy to see that these diagrams depend only on the 
position $n_1$ of the first crossing 
(i.e. the second occurrence of the particle $1$) and are given by
\begin{equation}
b_2(n_1,n)=\frac{n_1-2}{2^{n-2}D}\frac{\rho^{n-1}(-\lambda)^n}{\kappa^{n-1}}
\int_{\mathbb{R}^{D}} {d\bq\over (2\pi)^D} {\tilde V}(\bq)^n.
\label{coefn1}
\end{equation}

Hence, taking into account all the multiplicities, 
the contribution of all the diagrams with $n>2$ lines is 
\begin{equation}
\begin{split}
b(n)&=b_1(n)+\sum_{n_1=2}^{n-1} 2^{n-n_1-1} b_2(n_1,n)\\
&=\frac{1-2^{1-n}}{D}\frac{\rho^{n-1}(-\lambda)^n}{\kappa^{n-1}}
\int_{\mathbb{R}^{D}} {d\bq\over (2\pi)^D} {\tilde V}(\bq)^n.
\end{split}
\end{equation}
At $n=2$  the above agrees  the result of Eq. (\ref{eqnE2}), and hence the
above formula is valid for all $n\geq2$.
Performing the sum we find the one-loop contributions to be 
\begin{widetext}
\begin{equation}
\kappa_e^{(\text{one-loop})}-\kappa=-\sum_n b(n)= 
-\frac{\rho\lambda^2}{D\kappa} \int_{\mathbb{R}^{D}} \frac{d\bq}{
(2\pi)^D} {\tilde V}({\bf
q})^2\left(\frac{1}{1+\frac{\rho\lambda}{\kappa}{\tilde V}(\bq)}- 
{1\over 2}{1\over{1+\frac{\rho\lambda}{2\kappa}{\tilde V}(\bq)}}\right).
\label{eq1loop}
\end{equation}
\end{widetext}
The above may be conveniently rewritten in terms of $c = \rho\lambda/\kappa$
\begin{eqnarray}
\kappa_e^{(\text{one-loop})}&=& \kappa\left[1 + {1\over \rho D}
\left(g(c) - 2g({c\over 2})\right)\right] \\
&=&\kappa\left[1 + {\beta\over c D}
\left(g(c) - 2g({c\over 2})\right)\right] \label{koneloop}
\end{eqnarray}
where
\begin{equation}
g(c) = \int_{\mathbb{R}^{D}} \frac{d\bq}{
(2\pi)^D} \frac{c{\tilde V}({\bf
q})}{1+c{\tilde V}(\bq)}
\end{equation} 
From the previous discussions the  corrections to Eq. (\ref{koneloop})
of the form
\begin{equation}
\kappa_e = \kappa_e^{(\text{one-loop})} + \beta^2 s_2(c) + \beta^3 s_3(c)
\cdots
\end{equation}

In the particular case of the potential given by Eq. (\ref{potgauss}), we
find that in $D$ dimensions
\begin{equation}
g(c) = {1\over (2\pi)^{{D\over 2}} \Gamma({D\over 2})}\int_0^\infty
du\ {c \exp(-u) u^{{D\over 2}-1}\over 1 + c \exp(-u)}
\end{equation}
In the case $D=2$, we obtain
\begin{equation}
\kappa_e^{(\text{one-loop})}=\kappa+\frac{\kappa}{4\pi\rho}\left[\ln\left(1+
\rho\beta\right)-
2\ln\left(1 + {\rho\beta\over2}\right)\right].
\end{equation}
However, this expression becomes negative when $\beta$ is large, although this
is outside the range where the approximation involved here is valid. We also
see that $\kappa_e$ is a non-monotonic function of $\rho$, having a minimum 
value at some value $\rho_c$ but increasing up to $\kappa$ again on taking
$\rho$ very large - this is a consequence of the use of a soft potential,
at high densities the particles all overlap but the energy change in moving 
to overlap with one particle rather than another is zero. The 
effective potential seen by the particles is almost flat and hence there
is only a small effect on the particle diffusion.

\subsection{One-loop renormalization-group}
We have seen in the precedent section that  the one-loop calculation of
$\kappa_e$ predicts  that the diffusion constant can vanish at finite
temperature. The same calculation predicts a vanishing diffusion 
constant for the problem of diffusion in a Gaussian random potential
with short range correlations. This transition can be shown to be
absent in finite dimensions via exact results in one and two dimensions,
numerical simulations and general arguments based on the fact that
the system has a finite correlation length. 
In order to go beyond the simple one-loop contribution, one can look at how
the effect of the interactions on the diffusion constant propagates from
short to large length scales. 
This can be achieved by the renormalization-group
method. This approach has been proven to be very
accurate in the calculation of the effective diffusivity in random
media and removes the fictitious vanishing of $\kappa_e$
predicted by simple perturbation theory. 
Indeed, in dimension three the one-loop renormalization
group analysis provides a very good quantitative approximation 
of the effective diffusivity of a
particle in a Gaussian random potential 
\cite{dean94,dc,dean95a,dean95b,dean96},
 and the exact result in dimensions one and two \cite{dean97}.
We therefore apply the same technique to the problem of interacting
particles studied here. 
The potential $\phi$ is decomposed into a long  and short
scale components
\begin{eqnarray}
\phi_{<}(\bx) = 
\int_{|q|<\Lambda} {d\bq\over (2\pi)^{ND}} {\tilde \phi}(\bq) 
\exp(i\bq \cdot \bx) \nonumber \\
\phi_>(\bx) = 
\int_{|q|>\Lambda} {d\bq\over (2\pi)^{ND}} {\tilde \phi}(\bq) 
\exp(i\bq \cdot \bx)
\end{eqnarray}
where $\Lambda$ is a running cut-off.
One then integrates out the high momentum component $\phi_>$
perturbatively to find an effective theory on length scales
greater than $1/\Lambda$. One then makes a self similarity ansatz which
means only keeping
interactions that were in the original problem and hence the 
only parameters which change are $\kappa$ and $\lambda$
(as well as the potential which just has the higher Fourier
modes removed). One therefore has a running diffusion constant
$\kappa(\Lambda)$ and a running coupling to the gradient field 
$\lambda(\Lambda)$. The effective diffusion constant is then given
by $\kappa_e =\kappa(0)$. 
The flows of these couplings can be
computed from  the one-loop diagrams. In addition it
can be shown by general arguments \cite{dean97} that
\begin{equation}
{\kappa(\Lambda)\over \lambda(\Lambda)}={\kappa\over \lambda}
= T,
\label{eqnfdt}
\end{equation}
that is to say that the Einstein relation or Fluctuation 
dissipation relation is satisfied by the renormalized theory
at each step of the renormalization. This renormalization
is only valid for the low-momentum component of the remaining
drift  $\nabla\phi_<$ and a possible improvement to the 
calculation here would be to functionally renormalize the 
field $\phi_<$, which amounts to introducing new interactions generated by
the renormalization procedure (this approach has been applied
to diffusion in an incompressible quenched Gaussian velocity field
\cite{dean2001}) .
Using the one-loop diagrams calculated in the previous section
we find the flow equation for $\kappa(\Lambda)$ is
\begin{widetext}
\begin{equation}
\frac{d\kappa^{(\text{one-loop})}(\Lambda)}{d\Lambda}=-\frac{\rho
\lambda(\Lambda)^2}{D\kappa(\Lambda)(2\pi)^D}S_{D-1} \Lambda^{D-1}
{\tilde V}(\Lambda)^2\left(\frac{1}{1+\frac{\rho\lambda(\Lambda)}{\kappa(\Lambda)}{\tilde V}(\Lambda)}- 
{1\over 2}\frac{1}{1+\frac{\rho\lambda(\Lambda)}{2\kappa(\Lambda)}{\tilde
V}(\Lambda)}\right),
\label{dkdl}
\end{equation}
\end{widetext}
where $S_{D-1}$ is the area of the unit sphere of $\mathbb{R}^D$. Inserting 
(\ref{eqnfdt}) into Eq. (\ref{dkdl}) one gets
\begin{widetext}
\begin{equation}
\frac{d\kappa^{(\text{one-loop})}(\Lambda)}{d\Lambda}=-\frac{\rho
\lambda(\Lambda)}{DT(2\pi)^D}S_{D-1} \Lambda^{D-1}
{\tilde V}(\Lambda)^2\left(\frac{1}{1+\frac{\rho}{T}{\tilde V}(\Lambda)}- 
{1\over 2}\frac{1}{1+\frac{\rho}{2T}{\tilde
V}(\Lambda)}\right).
\end{equation}
\end{widetext}
Integrating out $\Lambda$ down from $\infty$ to $0$
and using the initial conditions $(\kappa(\infty), \lambda(\infty))
= (\kappa, \lambda)$, we find the final expression for
the effective diffusion constant is simply
\begin{equation}
\kappa^{(\text{one-loop RG})}_e=\kappa\exp\left({\frac{\kappa^{(\text{one-loop})}_e}{\kappa}-1}\right).
\end{equation}
Now looking at $\kappa^{(\text{one-loop})}_e$ given in
Eq. (\ref{eq1loop}), the value of the self-diffusion constant given by
the self-similarity ansatz leads to several comments:
\begin{itemize}
\item The temperature-density dependence of $\kappa_e/\kappa$ is
  of the form $ h({\rho\over T})/T$. Hence the
  phase diagram in the $(T,\rho)$ plane obtained from it consists of
  regions separated by straight lines crossing at the point
  $(0,0)$. However, the Langevin approach studied here is in most 
  physical situations a coarse-grained approach, in which case the
  potential $V$ has to be replaced by an effective potential
  $V_{\text{eff}}(\bx)= V_{\text{eff}}(\bx, c,\rho)$ 
which depends on the temperature
 and possibly the density. In this case, the phase diagram will be more
  complicated.
\item If the Fourier transform ${\tilde V}(q)$ of the potential has an absolute
  minimum {\it of negative value} at some non-zero value $q^*$, then the
  diffusion constant goes to zero when the temperature reaches
  $T^*=-\rho\,(\tilde{V}(\bq^*)^{-1}$ from above. Keeping the density fixed
lowering $T$ amounts to increasing $c$. The integral
\begin{equation}
I = \int d\bq {c {\tilde V}(\bq)\over 1 + c {\tilde V}(\bq)}
\end{equation}
diverges at $c = c^* = \rho/T^*$, and near $c^*$ the denominator of the 
integrand behaves as 
\begin{eqnarray}
(1 &+& (c^*- \delta c)(V(q^*) + 
V_{,qq}(q^*){\delta q}^2/2 ))\nonumber \\
&=& (-\delta c V(q^*) + V_{,qq}(q^*){\delta q}^2/2),
\end{eqnarray} 
where we have assumed that
$V$ is twice  differentiable about $q^*$, {\em i.e.} the minimum is 
not a cusp. In this case the  relaxation time diverges at $T^*$ as 
\begin{equation}
\tau \sim  \exp\left( \frac{A}{\sqrt{T-T^*}}\right). \label{eqtas} 
\end{equation}
However one could conceive that $V$ is an
effective potential coming from a course grained approach to a
microscopic  model and that it may thus have
a dependence on $\rho$ or $T$. Alternatively one might argue that
diagrams of lower order in $\rho$ dress the effective interaction 
leading to and additional $\rho$ or $T$ dependence in $V$.
In this scenario, if we keep $\rho$ fixed, the
denominator of the  integrand
now behaves as  
\begin{eqnarray}
(1 &+& (c^*- \delta c)(V(q^*,c^*) -
V_{,c}(q^*,c^*)\delta c + V_{,cc}(q^*,c^*){\delta c}^2/2 \nonumber \\
&+& 
V_{,qq}(q^*,c^*){\delta q}^2/2 +  V_{,qc}(q^*,c^*)\delta c\delta q))
\nonumber \\ &\approx&
(\delta c/c^*  -
c^* V_{,c}(q^*,c^*)\delta c - V_{,qc}(q^*,c^*)\delta c\delta q))
\end{eqnarray}
By definition we have that $V_{,c}(q^*,c^*) < 0$ as the minimum of $V(q,c)$
arrives at $-1/c$ from above. Hence is this case the relaxation time diverges
as
\begin{equation}
\tau \sim  \exp\left( \frac{A}{T-T^*}\right), \label{eqtvf}
\end{equation}
which is the Vogel-Fulcher-Tammann law. From the above analysis we see that
if the dependence of $V(q,c)$ on $c$  is weak near $(q^*,c^*)$ then one would
expect to see a crossover between the behavior of $\tau$ between Eq. 
(\ref{eqtas}) for $|T-T^*|< 1$, to the behavior 
Eq. (\ref{eqtvf}) when  $|T-T^*| \ll 1$.
If $q^*=0$ then the behavior is 
different and depends on the spatial dimension $D$.   
Finally as mentioned in the previous section on the one-loop expansion, 
precisely  at $T=T^*$ one expects that $\kappa_e$ depends explicitly on $N$.
\item In the corresponding static approximation \cite{acsa} it was shown
that the  structure function is given by
\begin{equation}
S(q) = {1\over 1 + c {\tilde V}(q)}, \label{eqsq}
\end{equation}
although the authors of \cite{acsa} express $S$ in terms of the Mayer
function of the potential $V$, the result is the same at the order
of accuracy of the calculation. We see that the condition above for 
a dynamic transition, defined by the vanishing of $\kappa_e$, coincides
in the same approximation scheme by the appearance of a diverging
correlation length and a second order phase transition. However as
pointed out in \cite{acsa} this apparent second order transition may
be preceded by a first order freezing transition when the 
free energy of the ordered crystal phase is lower than that of the
free energy of the liquid phase described by the terms in the 
chain re-summed virial expansion. The analysis above thus may apply to
a supercooled liquid and the temperature $T^*$ is the limit of the
thermodynamic stability of the liquid phase. There has been some experimental
evidence of a diverging correlation length at the Vogel-Fulcher-Tammann
temperature $T^*$ from the study of the dielectric susceptibility of
supercooled liquids \cite{mena}, though these results seem at odds with
earlier numerical studies \cite{erna,dain} studying this question. 
 
\end{itemize}

\section{Comparison with  Monte Carlo simulations}

\begin{figure}[htbp]
\includegraphics[width=.45\textwidth]{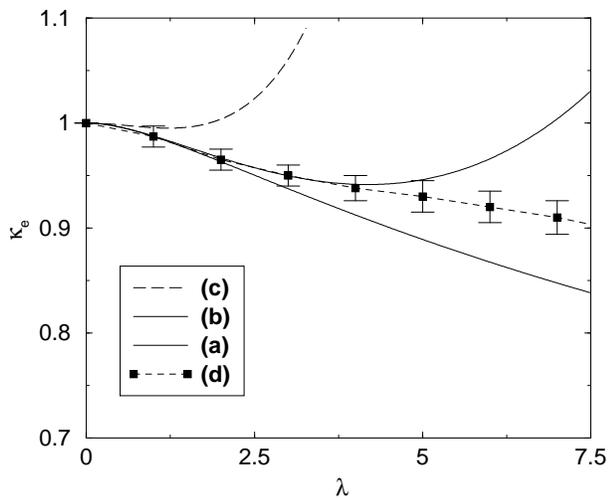}
\caption{Microscopic diffusion constant as a function of the interaction
from different approximation schemes (a,b,c), compared to the one measured
in Monte Carlo simulations: $\frac{1}{T}$-expansion from
Eq. (\ref{eqkappas}) and (\ref{eqkappas1} (a), one-loop
renormalization-group (b), one-loop renormalization-group $+$ first
two-loops diagram (c) and Monte Carlo simulations (d).}
\label{fig8}
\end{figure}

In order to test the accuracy of the different schemes explained above, 
we have carried
out Monte Carlo simulations of particles interacting via the potential
(\ref{potgauss}) in two dimensions. Here we have fixed the density at the
value $\rho=0.5$, and the number of particles at $N=10000$. 
We have
evaluated the macroscopic diffusion constant by using Eq. (\ref{defk}).
The result is shown on Fig. (\ref{fig8}) and compared to the evaluation from
Eq. (\ref{eqkappas}) and (\ref{eqkappas1}, for values of the interaction in the
range $[0,7.5]$. 

From  Fig. (\ref{fig8}) we see that 
the $\beta$-expansion is valid only at very small
$\beta$, however the one-loop RG result is much nearer the simulated
values of the diffusion constant over  quite a broad range. We expect that
a  RG analysis including more interactions, for example including
functional renormalization of the interaction $V$,
would provide better agreement. The two-loop
diagrams include the one indicated as type 1 in Fig. (\ref{fig5}). The
effect of the addition of this diagram to the one-loop RG calculation
is plotted in Fig. (\ref{fig8}) as well. The other diagrams of the
two-loop expansion are at least of order $\beta^4$. Hence, the good
agreement observed indicates that the two-loop calculation which includes
additional interactions  will improve the approximation to the 
diffusion constant. We note
that the value of the density used here ($\rho=0.5$) is not very
small, and the reasonably good agreement obtained is a hint that the
RG calculation catches the $\rho$ dependence of the diffusion
constant in the case of soft or bounded potentials.

\section{Conclusion}
We have developed a perturbative expansion of the self-diffusion
constant of Langevin interacting particles. The expansion is a 
weak coupling expansion which is suitable for soft or bounded interaction 
potentials. The perturbation expansion can be seen to be a double expansion
in $\beta$ and $\rho$, and we have shown how it can be dealt with 
diagrammatically. As an example of a partial re-summation of this
expansion,  a one-loop renormalization group analysis has been carried out 
and tested on a simple form of the interaction, and compared to numerical
simulations. It was found that this calculation gives
considerably better agreement than the straight expansion to
$O(\beta^3)$. In addition, in some cases the one-loop RG calculation 
predicts a divergence of the relaxation time  (or vanishing of the self 
diffusion constant) at some positive temperature. However we have seen  that
when such a dynamic transition occurs it is accompanied by a diverging  
correlation length in the statics when one uses equivalent approximations
in the statics and dynamics. The possible forms of the divergence
of the relaxation time have been discussed and it has been argued that
the Vogel-Fulcher-Tammann law emerges under relatively weak 
additional assumptions to the basic calculation carried out here.  
The two-loop calculation should provide a better quantitative 
approximation for the diffusion constant, but it would also 
indicate  the form or robustness of the divergence of the relaxation 
time as higher-loop contributions are taken into account, this calculation 
is in progress. Various other re-summation methods may be developed
on the systematic perturbation expansion expounded here, as we have 
mentioned earlier the series may be re-summed at order $\rho$ to 
recover existing low density results, one may use self consistent
perturbation theory and also explore renormalization group schemes
based of calculating the effect of integrating out the effect of a small
density of other particles rather than the Fourier modes of the interaction 
potential.

\end{document}